# Neutral test particle dynamics around the Bardeen-AdS black hole surrounded by quintessence dark energy

Jiayu Xie, Bing Tang[*]

*Department of Physics, Jishou University, Jishou 416000, China*

**ABSTRACT**

Dynamics of neutral test particles in the spacetime of a Bardeen-AdS black hole surrounded by quintessence dark energy is studied. First, we analyze the properties of the black hole and possible values of the monopole charge $\beta$ and quintessential parameters $a$ and $\omega_q$ that allows the existence of the event horizon. The effects of the parameters on the effective potential and the innermost stable circular orbit (ISCO) radius are also studied. For the neutral test particles motion, it is shown that as the quintessential parameters increase, the radius of ISCO be increased. We have analyzed the dynamical behaviors of the neutral test particles by applying techniques including Poincaré sections, power density and bifurcation diagram. It is shown that the presence of a quintessence parameter creates the chaotic phenomenon for the motion of neutral particle in a Bardeen-AdS black hole spacetime. The amplification of chaos typically occurs as the energy increases under appropriate circumstances.

## 1. Introduction

The study of particle motion in a black hole background helps to understand the nature of spacetime and to explain the spatial geometric structure around the black hole [1]. The event horizon of black holes, which acts as one way mem-brane, is one of the most fascinating boundaries gravity. The intimate relationship between the geometrical properties of spacetime horizons and the dynamics of particle motion near it has always been an interest to many researchers. The chaotic motion of particles seems to have a certain connection with the event horizon, one of the most important feature of chaos is that the tiny errors in the chaotic motion grow at an exponential rate, which causes the motion to completely deviate from the non-error state[2-4]. In last few decades, near horizon physics, at classical as well as quantum level, has been the active area of research [5-8]. There have been many attempts to study the particle dynamics in presence of event horizon for different kind of black hole systems where the black hole is either static spherically symmetric[9, 10] or rotating[5] or magnetized[8]. In this regard, circular geodesics provide the required information on spacetime geometry[11]. Similarly, the motion of test particles facilitates the observation of the gravitational fields of the objects experimentally and allows the comparison of the observed effects (such as light deflection, gravitational time delay, and perihelion shift) with the predictions of the effects.

[*] Corresponding author.
E-mail addresses: bingtangphy@jsu.edu.cn

In general relativity, the geodesics of the common Schwarzschild[10, 12], Reissner-Nordström[13], and Kerr[5, 14] spacetime (as well as all axisymmetric and static black-hole geometries) are integrable, and there is no chaotic behavior in the geodesic motion of the particles[8]. Because the geodesic equations of particles are variably fractional and the dynamical system is accretive, researchers have concluded that geodesic motion is not the best way to detect chaos arising from black holes. Martin's group found [15-17] that the motion of charged particles in magnetized Schwarzschild or Kerr black hole backgrounds is not accretive. Li found [8] that static motion of neutral and charged particles in the vicinity of the axisymmetric magnetized Ernst-Schwarzschild black holes may be chaotic, with no electromagnetic interaction of the neutral particles, and their chaotic behavior caused by the gravitational effect of an external magnetic field. Magnetic or electromagnetic fields can act as perturbations to the spacetime geometry, destroying its accretion and causing chaos[18]. Therefore, in order to ensure the non-accumulation of the particle's dynamical system and to further study its chaotic behavior, it is necessary to resort to some spacetime with a complex geometry or to introduce some additional interactions. In a previous study [19], we have investigated the effect of dark energy on the chaos of strings around the quintessential Bardeen-AdS black hole. Subsequently, the motion of neutral particles in the vicinity of a Bardeen-AdS black hole surrounded by quintessence dark energy is investigated to find out whether the action of dark energy can induce chaotic phenomena. To address these questions, we intend to calculate the energy and momentum corresponding to the particle motion in the innermost stable circular orbits (ISCO). We further aim to analyze how the quintessence parameter, magnetic charge effect the effective potential of neutral particles moving around black hole, the goal is to analyze the stability of orbits for a moving particle around black hole[20]. Then we investigate the chaotic phenomenon in motion of the particle in Bardeen-AdS black hole background surrounded by quintessence dark energy by the Poincaré section[21], power spectrum[22], and the bifurcation diagram[23].

We introduce the Bardeen-AdS spacetime surrounded by quintessence dark energy and discuss the effective potential and the innermost stable circular orbits at the equatorial plane in sect. 2. We survey the dynamics of generic orbits by using numerical methods in sect. 3. We present the main results and conclusions in sect. 4. Here, the constant of gravity $G$ and the speed of light $c$ take the geometrized units: $G=c=1$.

## 2. Spacetime Bardeen-AdS Black Holes Surrounded by Quintessence Dark Energy

In this section, an evolution model of neutral particles in a Bardeen-AdS spacetime geometry surrounded by quintessence dark energy is introduced. Then, the effective potential and innermost stable circular orbits (ISCO) at the equatorial plane are discussed.

The geometry of the spacetime metric around Bardeen-AdS surrounded by quintessence dark energy is in spherical coordinates, $x^\mu = (t,r,\theta,\phi)$ is given by

$$ds^2 = g_{\mu\nu}dx^\mu dx^\nu$$
$$= -f(r)dt^2 + \frac{dr^2}{f(r)} + r^2(d\theta^2 + \sin^2\theta d\phi^2) \quad (1)$$

where $r$ is the radial distance from the origin, and with the following gravitational

radially dependent the four-dimensional asymptotically AdS spacetime metric function

$$f(r) = 1 - \frac{2Mr^2}{(r^2+\beta^2)^{3/2}} + \frac{r^2}{l^2} - \frac{a}{r^{3\omega_q+1}} \qquad (2)$$

where $M$ is the total mass of the black hole, $\beta$ is a parameter of the Bardeen model and characterizes the monopole charge of a self-gravitating magnetic field described by a nonlinear electrodynamics [24, 25]. The energy density for quintessence has the form $\rho_q = -\frac{a}{2}\frac{3\omega_q}{r^{3(\omega_q+1)}}$, where $\omega_q$ is the quintessence state parameter that varies in the range $[-1, -1/3]$, and the integral constant $a$ is regarded as a positive normalization factor determined by the density of quintessence.

2.1 Dynamical model of neutral particles

The Lagrangian density for a neutral particle with mass $m$ is[26]

$$\begin{aligned}\mathcal{L} &= \frac{m}{2} g_{\mu\nu} \dot{x}^\mu \dot{x}^\nu \\ &= \frac{m}{2}(-f(r)\dot{t}^2 + f(r)^{-1}\dot{r}^2 + r^2\dot{\theta}^2 + r^2\sin^2\theta\dot{\phi}^2)\end{aligned} \qquad (3)$$

Conservative quantities of the motion can be easily found by using the Euler Lagrange equation. We define the covariant momentum

$$-E = p_t = \frac{\partial \mathcal{L}}{\partial \dot{t}}, \quad L = p_\phi = \frac{\partial \mathcal{L}}{\partial \dot{\phi}}, \qquad (4)$$

where the dot denotes differentiation with $\tau$, $E$ represents energy, $L$ is the angular momentum of the particle. And the conserved quantities of motion read

$$\dot{t} = \frac{E}{f(r)}, \quad \dot{\phi} = \frac{L}{r^2 \sin^2\theta}, \qquad (5)$$

Equations of motion for a test particle are then governed by the normalization condition

$$g_{\mu\nu}U^\mu U^\nu = \kappa, \qquad (6)$$

where $U^\mu = \dot{x}^\mu$ is the 4-velocity of the particle, $\kappa$ is 0 and $-1$ for massless and massive particles, respectively. Utilizing the normalization conditions in Eq.(6) taking into consideration Eqs.(4)-(5), we obtain the equations of motion in the separated and integrated form as

$$\dot{r}^2 = E^2 - f(r)(1 + r^2\dot{\theta}^2 + \frac{L^2}{r^2 \sin^2\theta}). \tag{7}$$

The geodesic motion of a particle is governed by the Hamiltonian:

$$H = \frac{1}{2} g^{\mu\nu} \frac{\partial S}{\partial x^\mu} \frac{\partial S}{\partial x^\nu}, \tag{8}$$

that is always identical to

$$H = -\frac{1}{2} m^2. \tag{9}$$

Hamilton's canonical equations of motion are given by

$$\dot{x}^\mu = \frac{\partial H}{\partial p_\mu}, \dot{p}_\mu = -\frac{\partial H}{\partial x^\mu}. \tag{10}$$

The following equations are obtained that describe the motion of the neutral particle.

$$\begin{aligned}
H &= -\frac{1}{2}(\frac{E^2}{f(r)} - p_r^2 f(r) - \frac{p_\theta^2}{r^2} - \frac{L^2}{r^2 \sin^2\theta}), \\
\dot{r} &= p_r f(r), \\
\dot{p}_r &= \frac{1}{2}[-\frac{E^2 f'}{f^2} - p_r^2 f' + \frac{2}{r^3}(p_\theta^2 + \frac{L^2}{\sin^2\theta})], \\
\dot{\theta} &= \frac{p_\theta}{r^2}, \\
\dot{p}_\theta &= \frac{L^2 \cos\theta}{r^2 \sin^3\theta}.
\end{aligned} \tag{11}$$

The event horizon of the spacetime around a black hole can be found by the condition $g_{tt}(r_h) = g^{rr}(r_h) = 0$. We present graphics for the event horizon radius of Bardeen-AdS black holes surrounding by quintessence dark energy. Fig. 1 provides the behavior of the dependence of the event horizon radius on the quintessential normalization factor $a$ for the different values of the magnetic charge $\beta$ and the state parameter $\omega_q$. It is seen from the figure that the Bardeen-AdS black holes have two solutions for the event horizon compared to the Schwarzschild-AdS black holes

($\beta = 0$). The inner horizon is the Cauchy horizon and it is unstable[27, 28]. We focus on how the outer horizon changes with these parameters. Other parameters are fixed, the outer horizon increases when the normalization factor $a$ increases. It grows rapidly when the values of $\omega_q$ is close to -1. On the contrary, as magnetic charge $\beta$ increases, the outer horizon decreases slightly. When $\omega_q$ decreases, the range of the positive quintessence normalization factor $a$ for the existence of the outer horizon is actually narrowed down. Moreover, there is no radius solution with $a$ in tiny value close to zero when $\beta$ increases to a specific value (for example, $\beta = 0.9$). This is consistent with the Ref.[24] that the Bardeen metric is a black hole with the condition $\beta^2 \leq \frac{16}{27}M^2$.

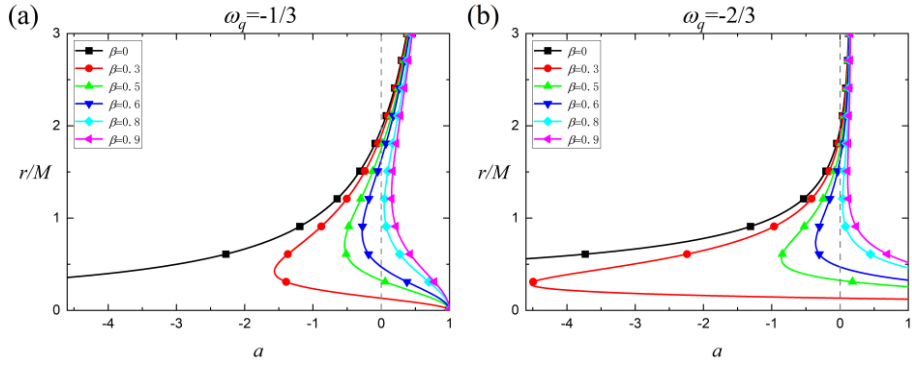

Fig. 1  The dependence of radius of the event horizon on the quintessential normalization factor $a$ for different values of the state parameter $\omega_q$ and the magnetic charge $\beta$.

2.2 Effective potential and innermost stable circular orbits

For the massive neutral particles the motion is governed by timelike geodesics of the spacetime, and the equations of motion can be found by using Eq.(7). Restricting motion of the particle to the plane, in which $\theta = \pi/2$ and $\dot{\theta} = 0$ in the following form:

$$\dot{r}^2 = E^2 - f(r)(1 + \frac{L^2}{r^2}). \qquad (12)$$

The equation of the radial motion can be expressed in the form

$$\dot{r}^2 = E^2 - V_{eff}, \qquad (13)$$

where $V_{eff}$ is the effective potential of the motion of neutral particles at the equatorial plane reads

$$V_{eff} = (1 - \frac{2Mr^2}{(r^2+\beta^2)^{3/2}} + \frac{r^2}{l^2} - \frac{a}{r^{3\omega_q+1}})(1+\frac{L^2}{r^2}). \qquad (14)$$

The properties of a neutral test particle near the field of a central black hole are mainly determined by the effective potential. Hence, near the quintessential Bardeen-AdS black holes, the impacts on the motion of a neutral particle can be derived based on how the effective potential depends on the magnetic charge $\beta$, the quintessence parameters $a$ and $\omega_q$. In Fig. 2, we exhibit various forms of the effective potential (14). It is seen from the figure that the radii of the extreme points become smaller as the magnetic charge $\beta$ increase and the particles can have orbits closer to the central black hole. The extreme points of the effective potential mean a test particle could move in circular orbits, noting that the orbit at the maximum (minimum) point is unstable (stable). These phenomena mean that the magnetic charge parameter induces the attractive effect on the motion of the neutral test particle. After the extreme point, in the absence of quintessential field ($a=0$), the $V_{eff}$ decreases relatively sharply as the radial coordinates increase. What is more, it is evident from the figure that $V_{eff}$ of the neutral particle decreases as the quintessence normalization factor $a$ increases. The decrease of state parameter $\omega_q$ also decreases the effective potential.

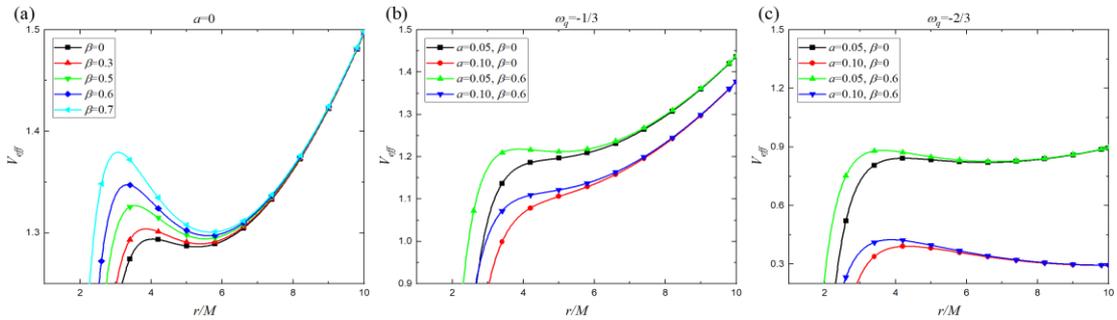

Fig. 2 The radial dependence of the effective potential of a neutral test particle around quintessential Bardeen-AdS black holes, the parameters are set as $L=4.5$, $l=15$.

Now applying standard conditions for the circular motion, namely no radial motion ($\dot{r} = 0$) and no forces in the radial direction ($\ddot{r} = 0$). Circular orbits correspond to the extreme values for effective potential are obtained if

$$E^2 = V_{eff}, V_{eff}' = 0, V_{eff}'' = 0, \tag{15}$$

where $'$ denotes the derivative with respect to the radius $r$, when $V_{eff}$ has its local minimum, $V_{eff}' = 0$, that is, a circular orbit exists. For $V_{eff}'' = 0$, the obtained stable circular orbit is the innermost stable circular orbit (ISCO). Then, we obtain

$$r_0 = \frac{3f(r_0)f'(r_0)}{2f'(r_0)^2 - f(r_0)f''(r_0)}. \tag{16}$$

The $r_0$ inscribing by the ISCO corresponding to effective potential. Traditionally, the stable circular orbits equation is defined by the standard condition $V_{eff}'' \geq 0$. In order to study the energy and angular momentum at ISCO corresponding to $r_0$, we have

$$E_{ISCO} = \frac{f(r_0)}{\sqrt{f(r_0) - r_0 f'(r_0)/2}}, \tag{17}$$

$$L_{ISCO} = r_0^{3/2} \sqrt{\frac{f'(r_0)}{2f(r_0) - rf'(r_0)}}. \tag{18}$$

Due to the complicated form of the exact expression for Eqs.(16)-(18), we only provide the numerical result for the ISCO radius to analyze the behavior of it. Fig. 3 provide the dependence of ISCO radius from the quintessential normalization factor $a$ at $\omega_q = -1/3$. We must perform some non-trivial algebraic operations in order to maintain the self-consistency of our result. It can be demonstrated that when $r \geq r_0$, the $E_{ISCO}$ and $L_{ISCO}$ remain positive. For the ISCO of the neutral test particle near the quintessential Bardeen-AdS black hole, the corresponding radius and angular momentum increase with the quintessence parameter $a$ when the magnetic charge $\beta$ is fixed. The presence of the magnetic charge $\beta$ causes the radius of the ISCO, $E_{ISCO}$ and $L_{ISCO}$ to decrease. However, when $\omega_q = -1/3$, the $E_{ISCO}$ has a minimum at a critical value of the quintessence parameter $a$. While less than the critical value, the

$E_{ISCO}$ decreases with $a$, and reversely it turns as increases with $a$. Furthermore, we present a parameter $\xi$ to describe the degree of deviation between the ISCO radius and the event horizon radius, which expresses as[29]

$$\xi = \frac{r_0 - r_h}{r_h}, \qquad (19)$$

We find that the increase of the quintessence normalization factor $a$ causes a decrease in the $\xi$, as shown in the Fig. 3 (b). In contrast, when $\omega_q = -2/3$, $\xi$ increases with $a$, as shown in the Fig. 4 (b).

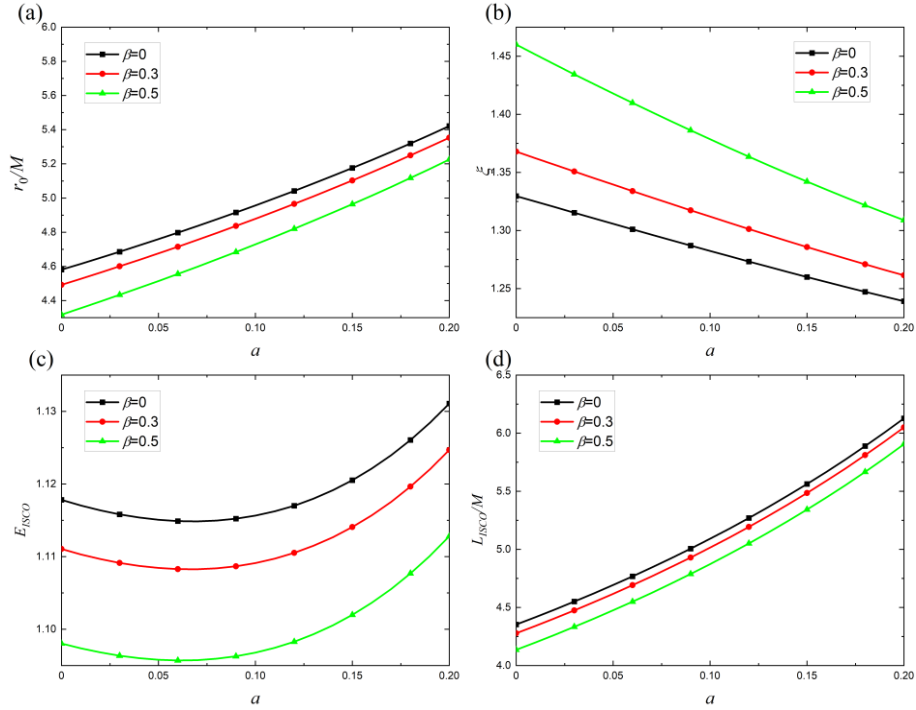

Fig. 3 Variation of the particle's energy per unit mass $E_{ISCO}$, its angular momentum $L_{ISCO}/M$ and its radial distance $r_0/M$ with the normalization factor $a$ for ISCO at $\omega_q = -1/3$. The panel (b) shows the corresponding deviation of ISCO from the event horizon with respect to $a$.

Fig. 4(a) provide the dependence of ISCO radius from the quintessential normalization factor $a$ at $\omega_q = -2/3$. Fig. 4(a) shows two possible orbits for the same values of parameter $a$, which is induced by the fact that the effective potential has two minima at $\omega_q = -2/3$. Further, the parameter $a$ has a maximum limit when the ISCO radius has only one solution. In this case, we only take the orbit closest to the event horizon and the corresponding $E_{ISCO}$ and $L_{ISCO}$ into consideration in Fig. 4. The term $\omega_q$ does not change the laws of the ISCO with quintessence normalization factor $a$, and does not affect the action of the magnetic charge $\beta$. When $\omega_q = -2/3$, the $E_{ISCO}$ is always decreases as the quintessence parameter $a$ increases. However, the angular momentum increases (quasi-)linearly with respect to increase of the quintessence parameter $a$ at $\omega_q = -1/3$ while the decreasing is power-law with $a$ at $\omega_q = -2/3$. In the following discussions, we apply numerical methods to study the dynamics of generic orbits.

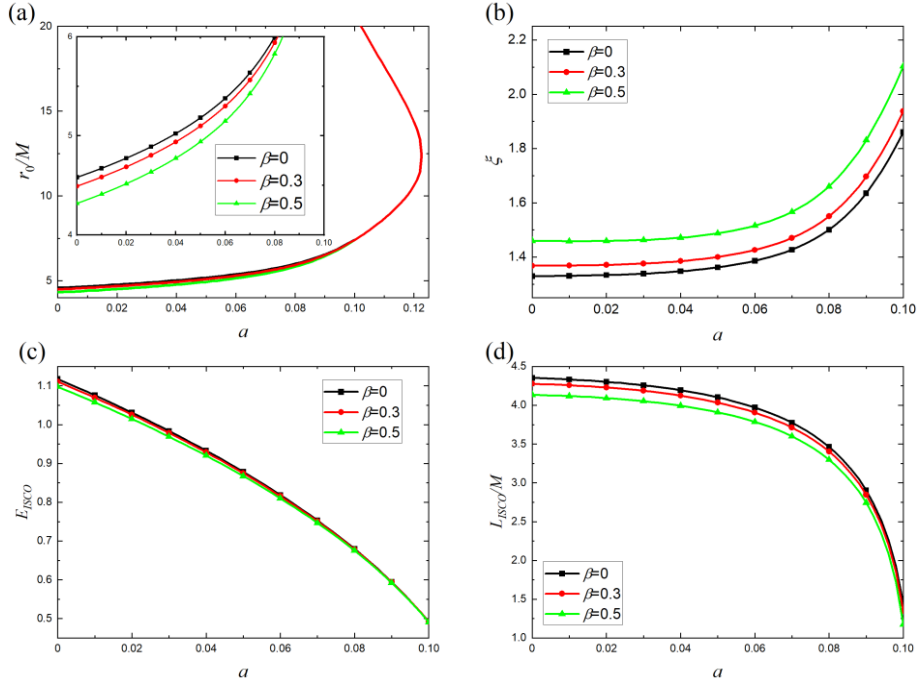

Fig. 4 Variation of the particle's energy per unit mass $E_{ISCO}$, its angular momentum $L_{ISCO}/M$ and its radial distance $r_0/M$ with the normalization factor $a$ for ISCO at $\omega_q = -2/3$. The panel (b) shows the corresponding deviation of ISCO from the event horizon with respect to $a$.

## 3. Investigations of orbital dynamics

3.1 Algorithms

To solve the equations of motion (11) numerically with high precision. The main reason is that the motion of a particle in chaotic region is very sensitive to initial value and the larger numerical errors may produce pseudo chaos which is not the real motion of particle. Here, we adopt to the corrected fifth-order Runge-Kutta method suggested in Refs.[12, 30-33], in which the velocities $(p_r, p_\theta)$ are corrected in integration and the numerical deviation is pulled back in a least-squares shortest path. As in refs. [33–36], the energy of the dynamical system is subjected to the constraint $H = -1/2$, which means that $H$ could be regarded as a conserved quantity. However, the numerical errors in the integral calculation could yield some deviations so that the numerical solution $(r, p_r, \theta, p_\theta)$ does not satisfy the constraint (9). In order to solve this problem, one can introduce a dimensionless parameter $\lambda$ to make a connection between the numerical velocities $(p_r, p_\theta)$ and the true value $(p_r^*, p_\theta^*)$ in the form of

$$p_r^* = \lambda p_r, \quad p_\theta^* = \lambda p_\theta. \tag{20}$$

The scale factor of velocity correction $\lambda$ can be chosen such that the constraint (9) is always satisfied. Inserting Eq. (20) into Eq.(11), one can find that the scale factor $\lambda$ in the quintessential Bardeen-AdS black hole spacetime is

$$\lambda = \sqrt{\frac{\frac{E^2}{f(r)} - \frac{L^2}{r^2 \sin^2 \theta} - 1}{p_r^2 f(r) + \frac{p_\theta^2}{r^2}}}. \tag{21}$$

In this way, the precision of the conserved quantity $\Delta H = H + \frac{1}{2} = 0$ in the system of Eq.(11) at every integration step can hold perfectly. In Fig. 5, we present the change of $\Delta H$ with time computed by the velocity correction method (RK5+Correction) in the quintessential Bardeen-AdS black hole spacetime. From Fig. 5, one can find that the value of $\Delta H$ is remained below $10^{-15}$ and then the error is controlled greatly, which displays sufficiently that this correction method is very powerful so that it can avoid the pseudo chaos caused by numerical errors.

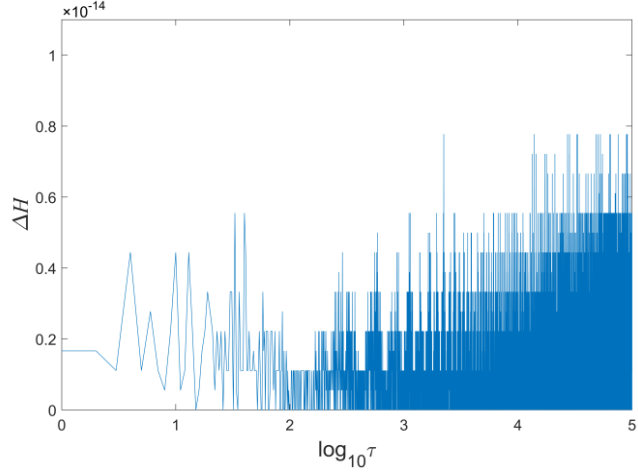

Fig. 5 Errors of $\Delta H$ with time computed by the velocity correction method (RK5+Correction) in the quintessential Bardeen-AdS black hole spacetime.

Here we consider Schwarzschild-AdS black hole ( $\beta=0, a=0$ ), Bardeen-AdS black hole ( $\beta \neq 0, a=0$ ) and quintessential Bardeen-AdS black hole ( $\beta \neq 0, a \neq 0$ ). Making use of the corrected fifth-order Runge-Kutta method, we obtain four numerical solution with a set of parameters $E=0.99, L=4.5, l=15, M=m=1$ and initial conditions $r(0)=10, p_r(0)=0, \theta(0)=\pi/2$, and then plot the changes of their radial coordinate $r$ with proper time $\tau$ in Fig. 6. It seems that these oscillations are regular, but their amplitudes and frequencies are difficult to be described by any definite pattern. We also plot the phase curve in the $(r, p_r)$ plane of the phase space for this trajectory in Fig. 7. It is shown that the solution with $\beta=0, a=0$ presented in Fig. 7(a) is a periodic solution and is not chaotic, which is explained by a fact that in the case $\beta=0, a=0$ is the usual Schwarzschild-AdS black hole spacetime in which the timelike geodesic equations are variable-separable and the chaos does not emerge in such an integrable dynamical system[34]. From the phase curve in the $(r, p_r)$ plane shown in Fig. 7(c)-(d), we find that the phase path becomes more complex and fulls densely a given region.

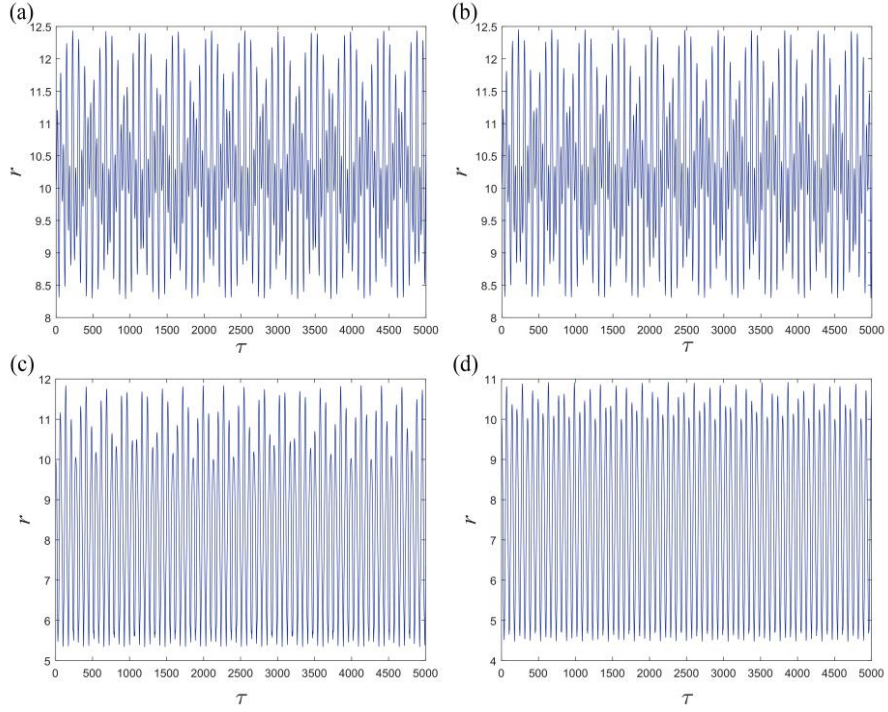

Fig. 6 The change of $r(\tau)$ with $\tau$ in the Bardeen-AdS black hole spacetime with a set of parameters $E = 0.99$, $L = 4.5$, $l = 15$, and initial conditions are $r(0) = 10$, $p_r(0) = 0$, $\theta(0) = \pi/2$. The panels (a), (b) are corresponding to the cases with the $\beta = 0$, $a = 0$ and $\beta = 0.3$, $a = 0$, respectively. The panels (c), (d) are corresponding to the cases with the $\omega_q = -1/3$, $\beta = 0.6$, $a = 0.12$ and $\omega_q = -2/3$, $\beta = 0.6$, $a = 0.06$, respectively.

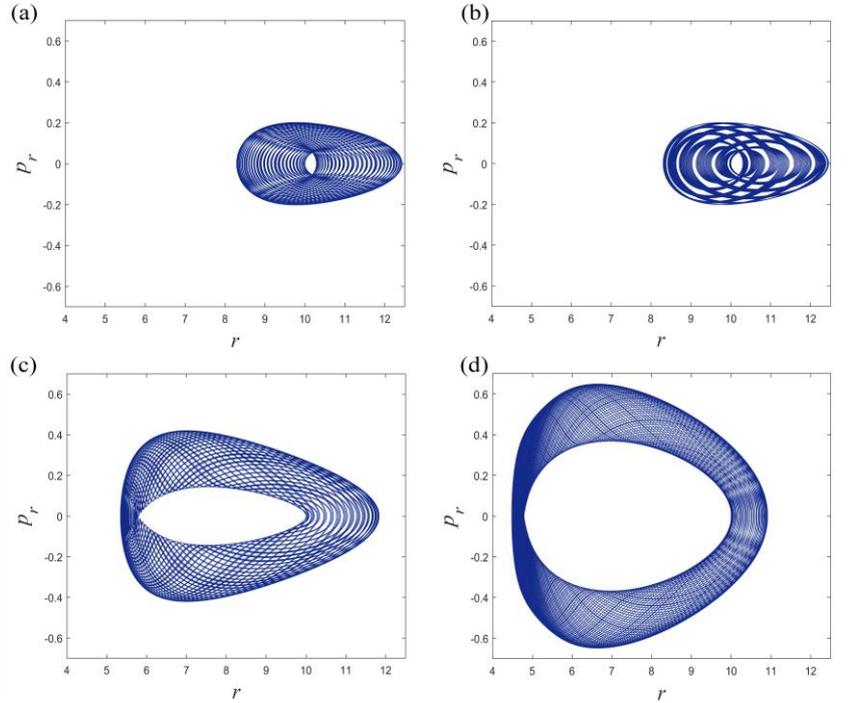

Fig. 7 Phase curves corresponding to the solutions plotted in Fig. 6, projected into the $(r, p_r)$ plane.

Thus, the presence of quintessential field make the motion of particle more complicated. It is shown that the phase path in the case with $a \neq 0$, the region of the phase path is enlarged and obviously approaches near the horizon, which means that the degree of disorder and non-integrability of the motion of particle increases with the quintessence dark energy.

3.2 Chaotic indicators

3.2.1 Bifurcation

The bifurcation diagram can tell us that the dependence of dynamical behaviors of system on the dynamical parameters, which could disclose the chaotic phenomenon in the system. In Fig. 8 and Fig. 9, we plot the bifurcation diagram of the radial coordinate $r$ with magnetic charge $\beta$ and normalization factor $a$ for the neutral particles moving in the Bardeen-AdS black hole spacetime with fixed $E = 0.999$, $L = 4.5$, $l = 15$. Here we chose the set of initial conditions are $r(0) = 10$, $p_r(0) = 0$, $\theta(0) = \pi/2$. For the chosen initial conditions plotted in Fig. 8(a), one can find that there is only multiple-periodic solution for the dynamical system (11), which means that the motions of particles are not chaotic in the case $a = 0$. For the case with the quintessential field in Fig. 8(b)-(c), one can find that with the decrease of the $\omega_q$, the $\beta$ of the chaotic motion (from 0.1 to 0.2) appears to be slightly larger. Fig. 8 Fig. 9(a)-(b), it is easy to find that there exist multiple-periodic, chaotic, attracted and escaped motion., After amplifying the region of Fig. 9(b) shown in Fig. 9(c), we find that they have the same shape and transformation. Thus the two cases of $\omega_q$, the increase of $a$ has a similar rule with a certain proportion of zooming. Moreover, we can find that with the increase of the magnetic charge $\beta$ and the quintessence normalization factor $a$, the motion of neutral particle transform between multi-periodic and chaotic motions, and the effects of the parameters $\beta$ and $a$ on the motion of the neutral particle are very complex, which are typical features of bifurcation diagram for the usual chaotic dynamical system, and then the chaotic motion occurs in these cases.

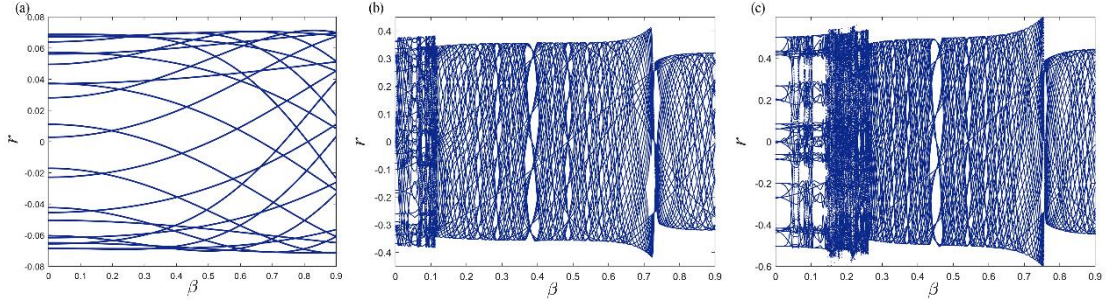

Fig. 8 The bifurcation diagram with magnetic charge $\beta$ in the quintessential Bardeen-AdS black holes spacetime. The set of parameters are (a) $a = 0$, (b) $\omega_q = -1/3$, $a = 0.1235$, (c) $\omega_q = -2/3$, $a = 0.0598$, respectively. The initial conditions are $r(0) = 10$, $p_r(0) = 0$, $\theta(0) = \pi/2$.

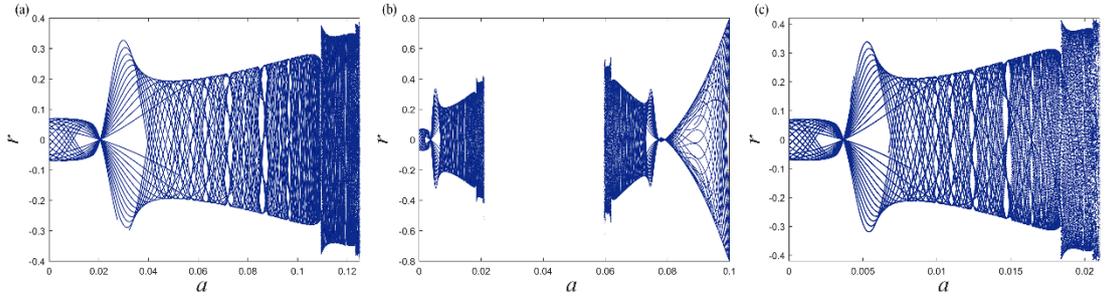

Fig. 9 The bifurcation diagram with quintessence normalization factor $a$ in the quintessential Bardeen-AdS black holes spacetime. The set of parameters are (a) $\omega_q = -1/3$, $\beta = 0.1$, (b) $\omega_q = -2/3$, $\beta = 0.1$. The panel (c) is a partial enlargement of (b). The set of initial conditions are $r(0) = 10$, $p_r(0) = 0$, $\theta(0) = \pi/2$.

3.2.2 Poincaré Section

The Poincaré section, which depicts orbits on a phase-space hypersurface with constant Hamiltonian, is a very useful tool for phase space visualization. A surface that is a transversal 2-dimensional surface where the Hamiltonian flow does not reside in the tangent plane to the surface at any point can be arbitrarily chosen within this hypersurface. A surface of section that is Poincaré section is what is known as such a hypersurface. It is the point in phase space where the trajectory of the particle intersects with a specific hypersurface that runs transverse to the trajectory. According to the intersection point distribution in Poincaré section, the motions of particles are classified as three kinds for a dynamical system. The periodic motions and the quasi-periodic motions correspond to a finite number of points and a series of close curves in the Poincaré section, respectively. Solutions for chaotic motion match unique groupings of scattered points with complex boundaries[21]. In Fig. 10 and Fig. 11, we show the change of the Poincaré sections in the plane $(r, p_r)$ with different quintessence parameters $a$ at $\omega_q = -1/3$ and $\omega_q = -2/3$. The section is defined by the condition $\theta = \pi/2$. The magnetic charge $\beta$ is fixed as $\beta = 0.1$. From Fig. 10, one can find that for $a = 0.12$ the phase path of the neutral particle is a quasi-periodic Kolmogorov–Arnold–Moser (KAM) tori[35] and the corresponding motion is regular under the effects of quintessential field. Moreover, we find that the KAM tori deforms with the increase of $a$. Since a regular orbit moves on a torus in the phase space and the corresponding curve in Poincaré sections is a cross section of the torus. Especially, as $a = 0.0605$, there is a chain of islands which is composed of seven secondary KAM tori belonging to the same trajectory. With the quintessence normalization factor $a$ increasing, the chain of islands are joined together and become a small KAM tori. This means that trajectory of the neutral particle in quintessential field is regular as $a > 0.06$.

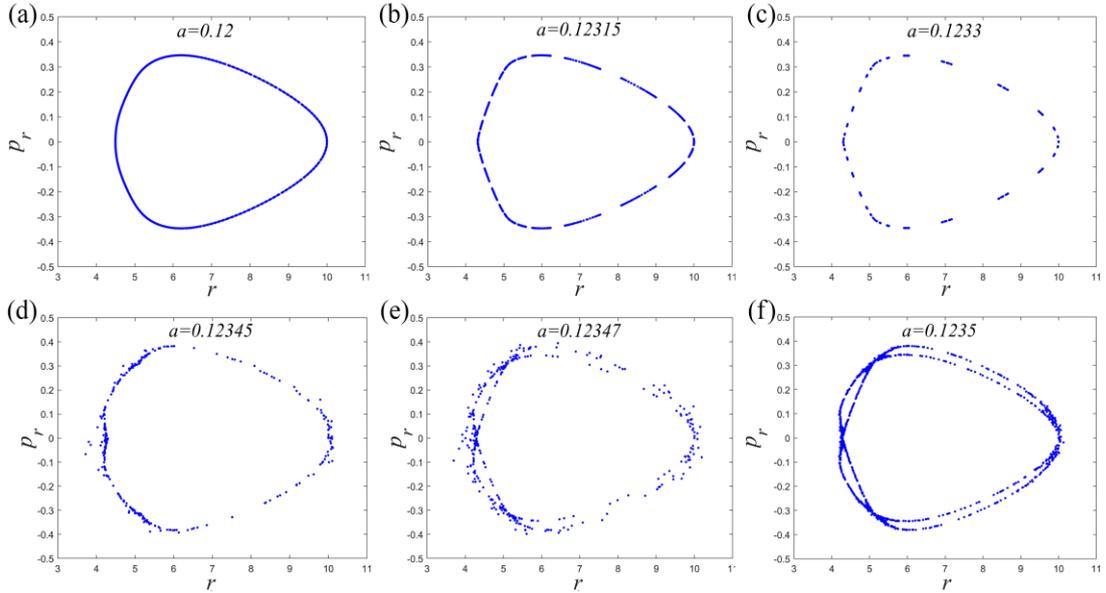

Fig. 10 The Poincaré surface of section ($\theta = \pi/2$) with different quintessence normalization factor $a$ and fixed magnetic charge $\beta = 0.1$ at $\omega_q = -1/3$. The parameters is fixed are $\beta = 0.1$, $E = 0.999$, $L = 4.5$ and $l = 15$.

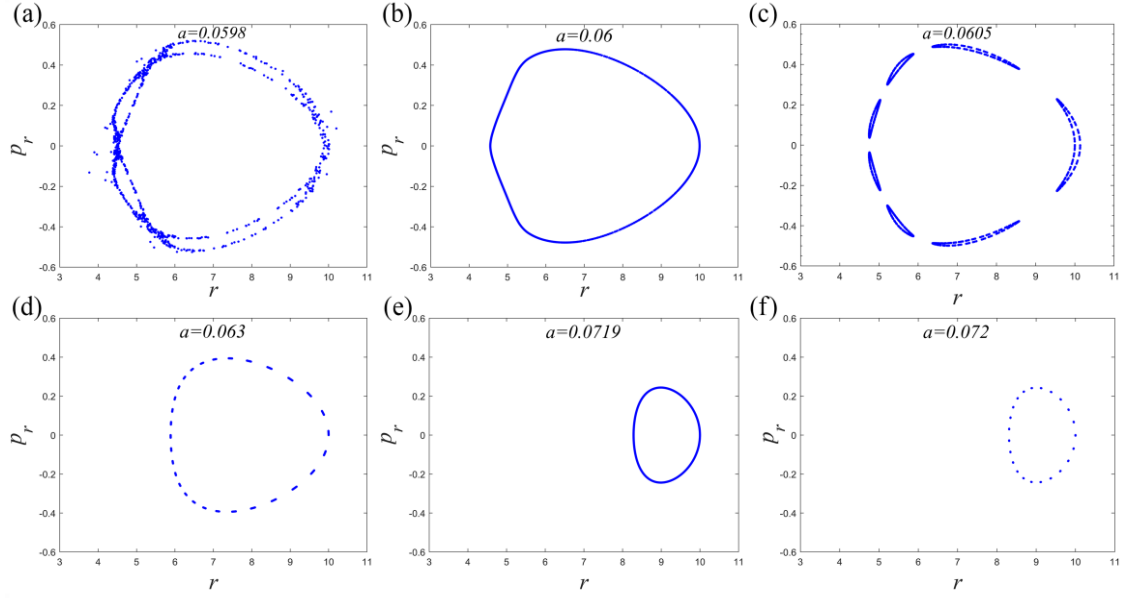

Fig. 11 The Poincaré surface of section ($\theta = \pi/2$) with different quintessence normalization factor $a$ and fixed magnetic charge $\beta = 0.2$ at $\omega_q = -2/3$. The parameters is fixed are $\beta = 0.2$, $E = 0.999$, $L = 4.5$ and $l = 15$.

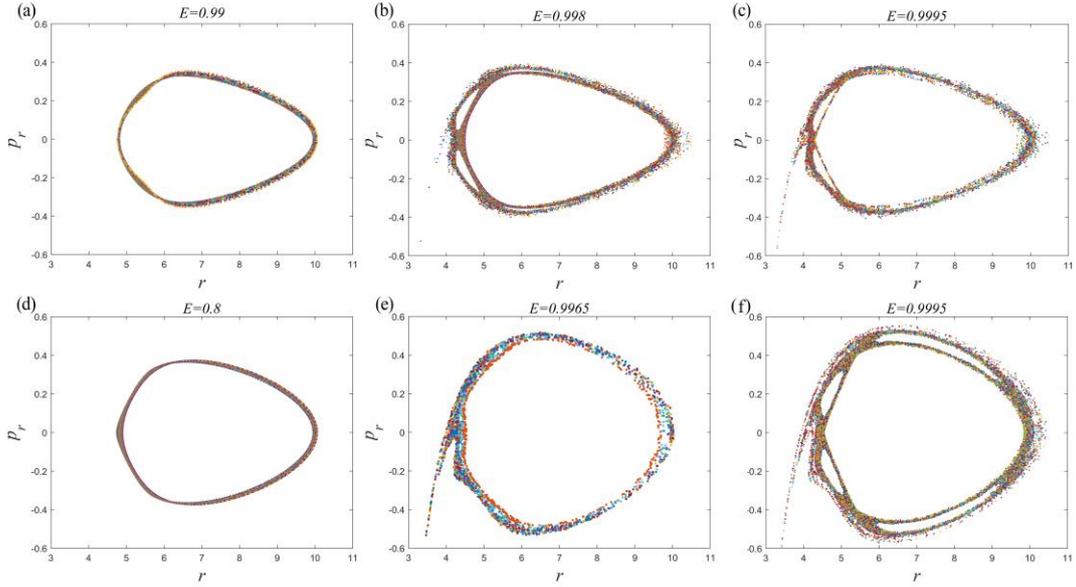

Fig. 12 The Poincaré surface of section ($\theta = \pi/2$) at different energies for the Bardeen-AdS black hole surrounded by quintessence. The first row is for $\omega_q = -1/3$ case, the other parameters is set as $\beta = 0.1$, $a = 0.1235$. The second row is for $\omega_q = -2/3$ case, the other parameters is set as $\beta = 0.2$, $a = 0.0598$.

In Fig. 12 we show the Poincaré section of the particle trajectory for the quintessential Bardeen-AdS black hole projected over the $(r, p_r)$ plane for different energies. We have considered the energies $E = 0.99, 0.998, 0.9995$ at $\omega_q = -1/3$ and $E = 0.8$, $0.991$, $0.9995$ at $\omega_q = -2/3$ as indicated in the plots. Different colors in the figures indicate the trajectory of the particles for different initial conditions. For low energy $E = 0.99$ and $E = 0.75$, with energy close to the vicinity of the minima of the effective potential, the Poincaré section exhibits the regular KAM tori. Therefore, the motion of orbits are quasi-periodic and regular. The regular motion is related to the regions in the close to those of the minima, This motion is supposed to be related to the observable quasiperiodic oscillations of radiation generated by regularly moving particles[8, 36, 37]. When the magnetic charge $\beta = 0.1$ is fixed while energy increases, such as $E = 0.998$, some of the orbits is no longer a torus and has a small number of randomly distributed points in Fig. 12(b), which indicates the presence of chaos. As energy further increases, such as $E = 0.9995$, the KAM tori is break and the $(r, p_r)$ plane is filled with the scattered points in Fig. 12(c), indicating that the extent of chaos is typically strengthened. In fact, chaos is established for energy growing sufficiently above the minimum of the effective potential. Additionally, as the total energy of the system is increased, the KAM tori starts getting distorted and appears to be attracted to the black hole horizon, which is the feature that supports the chaotic nature of the particle trajectory related to near the event horizon[38].

3.2.3 Power Spectral Density

In order to have an instinctual idea about the dynamical behavior of our composite system surrounding by dark energy and to understand the collective impact of horizon in the particle trajectories, we systematically investigate the evolution of the position coordinate, $x(\tau)$ with respect to the affine parameter. To get an extensive idea of the trajectory of the system and the deeper understanding of the onset chaotic behavior of the system, the power spectral density (PSD) is analyzed which is defined as[22, 39]

$$PSD = \frac{1}{2\pi \mathcal{N}} |x(\mathcal{N}, f, \Delta \tau)|^2, \tag{22}$$

where $x(\mathcal{N}, f, \Delta \tau)$ is the discrete Fourier transform of $x(\tau)$ evaluated at $\tau = k\Delta \tau$

($k = 0, 1, ..., \mathcal{N}$) and $\mathcal{N}$ is the length of the discrete affine parameter series.

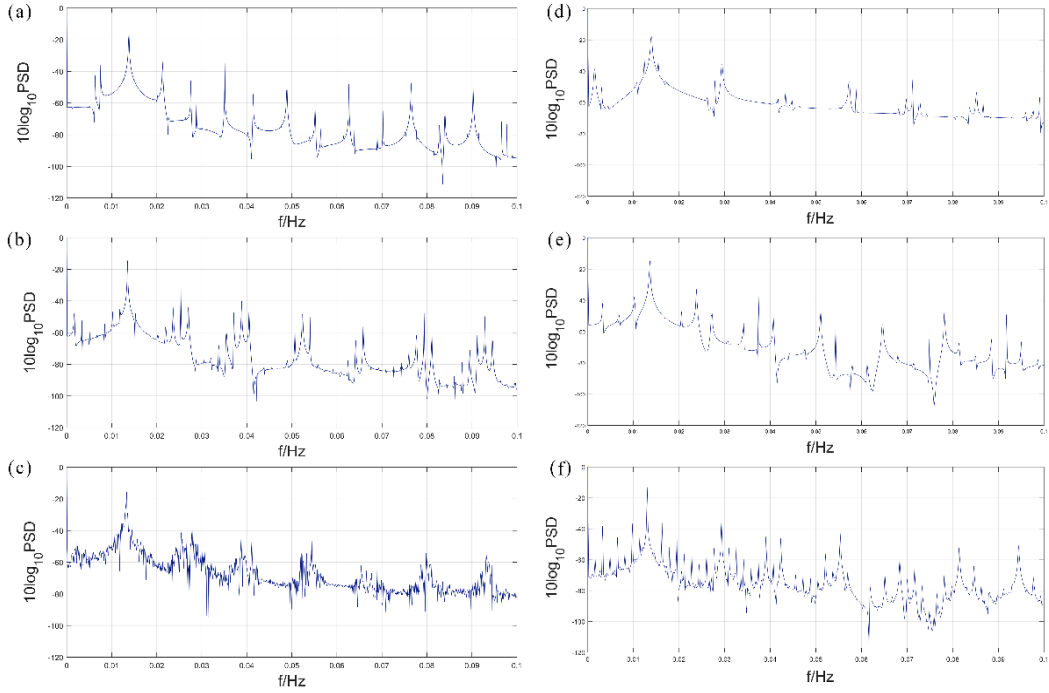

Fig. 13 The power spectrum for a neutral particle with $L = 4.5$, $l = 15$ for different values of quintessence parameter $a$ at $\omega_q = -1/3$, $E = 0.999$, namely (a) $a = 0.12$, (b) $a = 0.12315$, (c) $a = 0.12345$. And for different values of energy $E$ at $\omega_q = -1/3$, $a = 0.1235$, namely (d) $E = 0.99$, (e) $E = 0.998$, (f) $E = 0.9995$.

In Fig. 13(a)-(c), we present the PSD diagrams for the massive particle near the Bardeen-AdS black hole in a quintessential field moving with a constant total energy of the system ($E = 0.999$) but for different values of quintessence parameter $a$. The numerical simulations are done in the similar process as Fig. 10 keeping all the values of the parameters same. From the figure it can be seen that from $a = 0.12315$ onwards more frequencies start populating the spectrum. At $a = 0.1235$, the frequencies are highly populated in Fig. 13(c), which indicates the existence of chaos. Next we plot the PSD diagrams for different values of the total energy of the system (see Fig. 13(d)-(f)). The values of the other parameters are set as $L = 4.5$, $l = 15$, $\omega_q = -1/3$, $a = 0.1235$, and the initial conditions are $r(0) = 10$, $p_r(0) = 0$, $\theta(0) = \pi/2$. For low energy value of the total energy of the system ($E = 0.99$) the system contains only a few number of peaks appear into the plot (see Fig. 13(d)). As the total energy of the system gets increased the appearance of more frequencies are noticed in Fig. 13(f) at high value of

$E = 0.9995$, the highly population of frequency spectrum indicates that the system is in the chaotic regime.

## 4. Conclusion

In this paper we study the dynamics of neutral particles around a Bardeen-AdS black hole surrounded by quintessence dark energy. We present firstly the equation of motion for the neutral particle in the quintessential Bardeen-AdS black hole spacetime. The effective potentials and the innermost stable circular orbits at the equatorial plane are discussed. The performed analysis of effective potentials have shown that the increase of the quintessence normalization parameter $a$ and the decrease of the state parameter $\omega_q$ lead to decrease of the effective potentials. From the study of ISCO, we find that the ISCO radius increases with the normalization parameter $a$ at both $\omega_q = -1/3$ and $\omega_q = -2/3$, while it decreases in the magnetic charge $\beta$. It is found that the energy and angular momentum in the $\omega_q = -2/3$ case have quite different changes in comparison to the $\omega_q = -1/3$ case. Then, we have studied the dynamical behaviors of the motion of neutral particles by numerical method. Through analyzing the bifurcation diagram, Poincaré sections, and the power spectrum of dynamical system, we confirm that the chaos exists in the geodesic motion of the particles in the quintessential Bardeen-AdS black hole spacetime. The neutral particle have no charge and electromagnetic force is absent, then the quintessential field acts as the gravitational effect and destroys the integrability of the original spacetime. As such, chaos possibly occurs. In $\omega_q = -1/3$ and $\omega_q = -2/3$ cases, chaos is strengthened typically as energy or dark energy density increases. When neutral particles move in the spacetime geometry, normalization $a$ play an important role in strengthening or suppressing the extent of chaos caused by the gravitational effect of the quintessential field. It would be of interest to generalize our study to the collisions of particles around the quintessential Bardeen black holes, such as magnetized, electrically neutral, and magnetically charged particles. Work in this direction will be reported in the future.